\documentclass{article}

\usepackage[T1]{fontenc}
\usepackage[english]{babel}
\usepackage{amssymb}
\usepackage{amsmath}
\usepackage{latexsym}
\usepackage{marvosym}
\usepackage{amsfonts}
\usepackage{graphicx}
\usepackage{color}
\usepackage{fullpage}
\usepackage[applemac]{inputenc}
\usepackage{eurosym}
\usepackage{enumerate}
\usepackage{fancyhdr}
\usepackage{bbold}
\usepackage{caption}

\usepackage[pdftex=true,hyperindex=true,colorlinks=true,linkcolor=blue,bookmarks
=true,citecolor=blue]{hyperref}

\newtheorem{definition}{D\'efinition}[section]
\newtheorem{prop1}[definition]{Proposition}
\newtheorem{cor}[definition]{Corollary}
\newtheorem{hypothese}[definition]{Assumption}
\newtheorem{hypotheses}[definition]{Assumptions}
\newtheorem{rem}[definition]{Remark}

\newcommand{\prob}{\mathbb{P}}

\newcommand{\e}{\varepsilon}

\newcommand{\ud}{\mathrm{d}}

\title{A new characterization of the jump rate for\\
piecewise-deterministic Markov processes\\
with discrete transitions}
\author{Romain Aza\"{\i}s$^{\,\dag}$ and Alexandre Genadot$^{\,\ddag}$}
\date{\small $^\dag$ Inria Nancy -- Grand Est, Team BIGS and Institut \'Elie Cartan de Lorraine, Nancy, France\\
$^\ddag$ Institut de Math\'ematiques de Bordeaux and Inria Bordeaux -- Sud Ouest, Team CQFD
}

\begin{document}

\maketitle

\noindent
\textbf{Abstract:} Piecewise-deterministic Markov processes form a general class of non-diffusion stochastic models that involve both deterministic trajectories and random jumps at random times. In this paper, we state a new characterization of the jump rate of such a process with discrete transitions. We deduce from this result a nonparametric technique for estimating this feature of interest. We state the uniform convergence in probability of the estimator. The methodology is illustrated on a numerical example.

\medskip


\section{Introduction}

This paper is devoted to the estimation of the jump rate of a piecewise-deterministic Markov process (PDMP in abbreviated form) whose kernel transition only charges a finite set of points. PDMP's have been introduced in the literature by Davis \cite{DAVIS} in the eighties as a general family of non-diffusion stochastic models. They form a class of continuous-time Markov processes involving deterministic motion punctuated by random jumps, which occur either when the trajectory hits the boundary of the state space or in a Poisson-like fashion with nonhomogeneous rate before. More precisely, the trajectory followed by a PDMP $(X(t))_{t\geq0}$ on a metric state space $E$ is defined from its three local characteristics $(\lambda,\mathcal{Q},\Phi)$:
\begin{itemize}
\item $\Phi:\mathbb{R}_+\times E\to E$ is the deterministic flow. It satisfies the semigroup property,
$$\forall\,x\in E,~\forall\,(t,s)\in\mathbb{R}_+^2,~\Phi(t+s|x) = \Phi(s|\Phi(t|x)) .$$
\item $\lambda:E\to\mathbb{R}_+$ is the jump rate.
\item $Q:\mathcal{B}(E)\times\overline{E}\to[0,1]$ is the transition kernel.
\end{itemize}
The deterministic exit time of $E$ following the flow $\Phi$ is defined by,
\begin{equation*}
\forall\,x\in E,~t^\star(x) = \inf\left\{ t>0~:~\Phi(t|x)\in\partial E\right\}.
\end{equation*}
We impose usual conditions on the main features of the process \cite[(24.8) Standard conditions]{DAVIS},
\begin{equation}\label{eq:standard}
\forall\,x\in E,~\exists\,\varepsilon>0,~\int_0^\varepsilon \lambda(\Phi(t|x))\,\ud t <\infty \qquad\text{and}\qquad\forall\,x\in\overline{E},~Q(E\setminus\{x\}|x) = 1.
\end{equation}

\noindent
As mentioned before we assume in this paper that the transition kernel $Q$ only charges a finite set of points,
$$\exists\,\mathcal{E}\subset E,\quad\#\mathcal{E}<\infty\quad\text{and}\quad\forall\,x\in E,~Q(\mathcal{E}|x) = 1 .$$

\noindent
Starting from any initial condition $X(0)=x$, the motion of $(X(t))_{t\geq0}$ may be described as follows. The distribution of the first jump time $T_1$ is given by,
\begin{equation}\label{eq:G:S}
\forall\,t\geq0,~\prob(T_1>t\,|\,X(0)=x) = G(t|x)\mathbb{1}_{\{t<t^\star(x)\}},
\end{equation}
where
\begin{equation}\label{eq:def:functionG}
G(t|x) = \exp\left[-\int_0^t \lambda(\Phi(s|x))\,\ud s\right].
\end{equation}
It means that the process jumps either when the flow $\Phi(\cdot|x)$ hits the boundary of the state space at time $t^\star(x)$ or in a Poisson-like fashion with rate $\lambda(\Phi(\cdot|x))$ before. Next the post-jump location $Z_1$ at time $T_1$ is defined in $\mathcal{E}$ through the transition kernel $Q$,
$$\forall\,y\in\mathcal{E},~\prob(Z_1=y|X(0)=x, T_1)=Q(\{y\}|\Phi(T_1|x)) . $$
The path between $0$ and the first jump time $T_1$ is given by,
$$\forall\,0\leq t\leq T_1,~X(t)=
\left\{
\begin{array}{cl}
\Phi(t|x)&\text{if $t<T_1$,}\\
Z_1 & \text{else.}
\end{array}
\right.
$$

\noindent
Now starting from the post-jump location $X({T_1})$, one chooses the next inter-jumping time $S_2=T_2-T_1$ and the future post-jump location $Z_2$ in a similar way as before, and so on. One obtains a strong Markov process with $(T_n)_{n\geq0}$ as the sequence of the jump times (where $T_0=0$ by convention) and $(Z_n)_{n\geq0}$ as the stochastic sequence of the post-jump locations. The inter-jumping times are defined by $S_0=0$ and, for any integer $n\geq1$, $S_n=T_n-T_{n-1}$. $(Z_n)_{n\geq0}$ forms a homogeneous Markov chain on $\mathcal{E}$.

\noindent
In this paper, both the post-jump locations $Z_n$ and the inter-jumping times $S_{n+1}$ starting from $Z_n$ are observed within a long time interval. We propose to estimate the jump rate function $\lambda$ under the aforementioned assumption that the transition kernel $Q$ only charges a finite set of points. No assumptions are made on the form of the underlying deterministic dynamic.

\noindent
Statistical inference of the jump rate of PDMP's in this general setting (weak assumptions on the dynamic and observation of both post-jump locations and inter-jumping times within a long time interval) has already been investigated in two very recent articles \cite{amg,krell}. Nevertheless, to the best of our knowledge, discrete transitions have never been studied from a statistical point of view. The nonparametric method provided in \cite{krell} exploits the particular form of the invariant measure of the post-jump locations when the deterministic flow is one-dimensional and increasing. The author of \cite{krell} shows the convergence in $\mathbb{L}^2$-norm of the proposed kernel estimator of the jump rate $\lambda$ with rate $n^{-s/(2s+1)}$ whenever $\lambda$ is in the H\"older space $\mathcal{H}^s(D)$, $D$ a compact subset of $\mathbb{R}_+^\ast$. In \cite{amg}, the authors consider multi-dimensional PDMP's and write the jump rate as the ratio of a conditional density over a conditional survival function. They propose to estimate both these functions separately and state the almost sure convergence as well as the asymptotic normality of the resulting jump rate estimator. The jump mechanisms considered in these papers are different: transitions admit a density with respect to the Lebesgue measure in \cite{amg} while they are deterministic in \cite{krell}. It is thus quite natural to investigate in the present article the framework of transitions on a discrete grid. Discrete transition kernels often model regime changes in different application contexts. For instance, in eletrophysiology, the mechanism of the voltage-gate ion channels, exhibiting discrete transitions between opening and closing, is responsible for the generation and propagation of action potentials in nerves and muscles. Such phenomena have been successfully modeled by PDMP's with discrete regime changes, see \cite{FL07,RTT16} and references therein. Discrete transition kernels are also used as good approximations of continuous transition probabilities obtained for example by
optimal quantization \cite{BdSD,br2012}.

\noindent
The authors of \cite{AzaisSJS14} have shown that the multiplicative intensity model developed by Aalen \cite{Aal78} in the seventies is not satisfied by PDMP's but only by a transformed version of the underlying process. They deduce a nonparametric method for estimating the conditional density of the inter-jumping times. The strategy followed in the present paper is different and complementary but exploits results published in \cite{AzaisSJS14}. Assuming that the jump rate of the aforementioned transformed version of the process may be decomposed in some basis of $\mathbb{L}^2_{[0,1]}$, we state a new characterization of the jump rate $\lambda$ (see Proposition \ref{prop:lambda}). The resulting estimator of this function is presented in \eqref{eq:def:estimateurlambda}. Our main result is the uniform convergence in probability of this estimator (see Proposition \ref{cor}). We would like to already emphasize that our estimation procedure does not require to know or observe the deterministic flow $\Phi$ as it is the case in \cite{amg,krell}, but only the deterministic exit time $t^\star$.

\noindent
The paper is organized as follows. Section \ref{s:2} is devoted to the new characterization of the jump rate $\lambda(x)$ stated from results of \cite{AzaisSJS14}. The estimation procedure and the main results of convergence are presented in Section \ref{s:3}. We provide a short simulation study to illustrate the good behavior of the estimate in Section \ref{s:4}. Finally, all the proofs have been deferred to Appendix \ref{s:app:proofs}.

\section{Characterization of the jump rate}
\label{s:2}

In light of \eqref{eq:G:S} and \eqref{eq:def:functionG}, conditionally on the event $\{Z_n=x\}$, $x\in\mathcal{E}$, the inter-jumping time $S_{n+1}$ is governed by the nonhomogeneous rate $\lambda(\Phi(\cdot|x))$ and right-censored by $t^\star(x)$. As a consequence, this context seems well-adapted to estimate the cumulative rate $\int_0^\cdot\lambda(\Phi(s|x))\ud s$ by a Nelson-Aalen type estimator (see \cite[IV.1. The Nelson-Aalen estimator]{Andersen}). This strategy is applicable if the multiplicative intensity model developed by Aalen \cite{Aal78} in the seventies is satisfied (see also \cite{Hjort90} on this topic). The multiplicative intensity model supposes that the product of the jump rate of interest with some predictable process is the stochastic intensity of some counting process whose number of jumps tends to infinity. From the methodology developed in \cite{ADG2013} for marked renewal processes, one may show that the one-jump counting-process $t\mapsto\mathbb{1}_{\{S_{n+1}\leq t\}}$ admits the function
$$t\mapsto\lambda(\Phi(t|Z_n)) \mathbb{1}_{\{S_{n+1}\geq t\}}$$
as stochastic intensity in the filtration $\sigma(Z_n)\vee\sigma(\mathbb{1}_{\{S_{n+1}\leq t\}})_{0\leq t<t^\star(Z_n)}$, which is a first step to identify the multiplicative intensity model. In other words, the process
$$t\mapsto\mathbb{1}_{\{S_{n+1}\leq t\}} - \int_0^t \lambda(\Phi(s|Z_n)) \mathbb{1}_{\{S_{n+1}\geq s\}}\ud s$$
is a $\sigma(Z_n)\vee\sigma(\mathbb{1}_{\{S_{n+1}\leq t\}})_{0\leq t<t^\star(Z_n)}$-continuous-time martingale. Nevertheless, the sum over $n$ of these processes is generally not a martingale because the future post-jump location $Z_{n+1}$ may contain some information on the previous inter-jumping time $S_{n+1}$ (see \cite[Remark 2.2]{AzaisSJS14}). Consequently the multiplicative intensity model is generally not satisfied in the filtration generated by the post-jump locations $Z_n$ and the one-jump counting processes $t\mapsto\mathbb{1}_{\{S_{n+1}\leq t\}}$ for estimating the conditional rate $\lambda(\Phi(\cdot|x))$ of PDMP's. However, it should be noted that this remark does not show that the Nelson-Aalen estimator is not consistent for PDMP's but only that the usual strategy to establish the convergence is not adequate.

\begin{rem}
Under the condition $\Phi(\cdot|x)=x$, the underlying process becomes a marked renewal process (without deterministic censorship), which applications (stock prices, repairable systems, etc.) and statistical inference have been often investigated in the literature. In this particular context, the multiplicative intensity model is satisfied for the estimation of the jump rate $\lambda$ (see for instance \cite[Theorem 3.1]{ADG2013}), and thus the Nelson-Aalen estimator of the cumulative rate has good asymptotic properties. As a consequence, the statistical analysis developed in this paper is not relevant in this special setting.
\end{rem}

\noindent
The strategy developed in \cite{AzaisSJS14} consists in showing that the double-marked renewal process $(Z_n,Z_{n+1},S_{n+1})_{n\geq0}$ satisfies the multiplicative intensity model with its own rate $\widetilde{\lambda}$. Indeed, for any couple $(x,y)\in\mathcal{E}^2$, the process $(M^n(t|x,y))_{0\leq t<t^\star(x)}$ defined by,
\begin{equation*}\label{eq:sm}
\forall\,0\leq t<t^\star(x),~M^n(t|x,y) = N^n(t|x,y) - \int_0^t \widetilde{\lambda}(s|x,y)Y^n(s|x,y)\ud s,
\end{equation*}
is a continuous-time martingale in some filtration (see \cite[Theorem 3.4]{AzaisSJS14} with sets $A=\{x\}$ and $B_k=\{y\}$ that satisfy conditions imposed in Subsection 2.2 Assumptions), where the processes $Y^n(\cdot|x,y)$ and $N^n(\cdot|x,y)$ are defined, for $0\leq t<t^\star(x)$, by
\begin{eqnarray*}
Y^n(t|x,y) &=& \sum_{i=0}^{n-1}\mathbb{1}_{\{Z_i=x,Z_{i+1}=y\}}\mathbb{1}_{\{S_{i+1}\geq t\}} ,\\
N^n(t|x,y) &=& \sum_{i=0}^{n-1}\mathbb{1}_{\{Z_i=x,Z_{i+1}=y\}}\mathbb{1}_{\{S_{i+1}\leq t\}} ,
\end{eqnarray*}
and the function $\widetilde{\lambda}$ is defined by,
\begin{equation}\label{eq:def:lambdatilde}
\forall\,0\leq t<t^\star(x),~\widetilde{\lambda}(t|x,y) = \frac{f(t|x) Q(\{y\}|\Phi(t|x))}{\int_t^{t^\star(x)} f(s|x) Q(\{y\}|\Phi(s|x))\ud s + G(t^\star(x)|x)Q(\{y\}|\Phi(t^\star(x)|x))},
\end{equation}
where $f(\cdot|x)$ is given by,
\begin{eqnarray}
\forall\,0\leq t<t^\star(x),~f(t|x) &=& -\partial_t G(t|x)\nonumber \\
	&=& \lambda(\Phi(t|x)) \exp\left[-\int_0^t \lambda(\Phi(s|x))\ud s\right] .\label{eq:Glambdaf}
\end{eqnarray}
Function $f(\cdot|x)$ is only the conditional probability density of $S_{n+1}$ conditionally on the event $\{Z_n=x\}$ on the interval $[0,t^\star(x))$. From now on, one may estimate the conditional cumulative rate $\widetilde{\Lambda}$ defined by
\begin{equation}\label{eq:Lambdatilde}
\widetilde{\Lambda}(\cdot|x,y) = \int_0^\cdot\widetilde{\lambda}(s|x,y)\ud s
\end{equation}
by the Nelson-Aalen estimator $\widehat{\widetilde{\Lambda}}^n(t|x,y)$ given by,
\begin{equation}\label{eq:naestimate}
\forall\,0\leq t<t^\star(x),~\widehat{\widetilde{\Lambda}}^n(t|x,y) = \int_0^t Y^n(s|x,y)^\ast\ud N^n(s|x,y),
\end{equation}
where, for $\alpha\in\mathbb{R}$, $\alpha^\ast = \alpha^{-1}$ if $\alpha\neq0$ and $\alpha^\ast=0$ if $\alpha=0$. The asymptotic properties of this estimator in the framework of PDMP's are given in \cite{AzaisSJS14}, the main convergence result being, under some ergodicity and regularity conditions,
$$\sup_{0\leq s\leq t}\left|\widehat{\widetilde{\Lambda}}^n(s|x,y) - \widetilde{\Lambda}(s|x,y)\right| \stackrel{\mathbb{P}}{\longrightarrow} 0,$$
for any $0<t<t^\star(x)$ (see \cite[Proposition 3.7]{AzaisSJS14}).
One may then smooth this Nelson-Aalen estimator by kernel methods inspired by Ramlau-Hansen technique \cite{Ram83} to get a nonparametric estimate of the modified jump rate. More precisely, the kernel estimator of $\widetilde{\lambda}(s|x,y)$, $0\leq s<t^\star(x)$, investigated in \cite{AzaisSJS14} is given by
$$\widehat{\widetilde{\lambda}}^n(s|x,y) = \frac{1}{b}\int_0^t K\left(\frac{s-u}{b}\right)\ud\widehat{\widetilde{\Lambda}}^n(u|x,y),$$
for some $t\in[s,t^\star(x))$, where $K$ is some kernel function and $b$ denotes the bandwidth parameter. The most interesting result of consistency is the uniform convergence in probability \cite[Proposition 2.7]{AzaisSJS14},
$$\sup_{r_1\leq s\leq r_2} \left|\widehat{\widetilde{\lambda}}^n(s|x,y)-\widetilde{\lambda}(s|x,y)\right| \stackrel{\mathbb{P}}{\longrightarrow} 0,$$
where $0<r_1<r_2<t^\star(x)$, for some (random) bandwidth sequence almost surely going to $0$.
The strategy followed in \cite{AzaisSJS14} consists in coming back to the function of interest $f(\cdot|x)$ by exploiting the relation \eqref{eq:def:lambdatilde}. Nevertheless, this only leads to a nonparametric estimation method of the conditional density of the inter-jumping times and not of the rate of interest.
In this paper, we propose a new approach that exploits \eqref{eq:def:lambdatilde} to get a direct estimator of $\lambda$.
This technique relies on the following characterization of $\lambda$ under a few regularity conditions.

\begin{hypothese}\label{hyp:lambdatilde} The Hilbert space $\mathbb{L}^2_{[0,1]}$ being endowed with its usual scalar product $\langle \cdot,\cdot\rangle$ and norm $\|\cdot\|_2$, there exists an orthonormal family $(B_p)_{p\geq0}$ in $\mathbb{L}^2_{[0,1]}$ such that, for any $(x,y)\in\mathcal{E}^2$,
$$\forall\,0\leq s\leq1,~\widetilde{\lambda}(s t^\star(x)|x,y) = \sum_{p\geq0} \langle B_p,\widetilde{\lambda}(\cdot\,t^\star(x)|x,y)\rangle B_p(s) .$$
\end{hypothese}

\begin{hypothese}\label{hyp:tstar}
The function $t^\star$ is upper-bounded on $\mathcal{E}$.
\end{hypothese}


\begin{prop1}\label{prop:lambda}
Under Assumptions \ref{hyp:lambdatilde} and \ref{hyp:tstar}, we have,
\begin{equation}\label{eq:charac:lambda}
\forall\,x\in\mathcal{E},~\lambda(x) = \sum_{p\geq0} B_p(0) \sum_{y\in\mathcal{E}}R(\{y\}|x) \theta_p(x,y),
\end{equation}
where $R$ is the transition distribution of the Markov chain $(Z_n)_{n\geq0}$, i.e. $R(\{y\}|x)=\mathbb{P}(Z_1=y|Z_0=x)$ for $(x,y)\in\mathcal{E}^2$, and,
\begin{equation}\label{eq:def:thetap}
\forall\,p\geq0,~\forall\,(x,y)\in\mathcal{E}^2,~\theta_p(x,y)=\int_0^1\widetilde{\lambda}\left(t^\star(x)\,u|x,y\right) B_p(u)\ud u.
\end{equation}
\end{prop1}

\section{Estimation procedure}
\label{s:3}

From the observation of the $n$ first post-jump locations $Z_i$ and inter-jumping times $S_{i+1}$, our estimation method consists in estimating the unknown parameters appearing in the characterization \eqref{eq:charac:lambda} of $\lambda$. The transition function is estimated by its empirical version,
$$\forall\,(x,y)\in\mathcal{E}^2,~\widehat{R}^n(\{y\}|x) = \frac{\sum_{i=0}^{n-1}\mathbb{1}_{\{Z_i=x,Z_{i+1}=y\}}}{\sum_{i=0}^{n-1}\mathbb{1}_{\{Z_i=x\}}},$$
while the $\theta_p(x,y)$'s given in \eqref{eq:def:thetap} are very intuitively estimated by,
$$\forall\,p\geq0,~\forall\,(x,y)\in\mathcal{E}^2,~\widehat{\theta}^n_p(x,y)=\frac{1}{t^\star(x)}\int_0^1B_p(u) \ud \widehat{\widetilde{\Lambda}}^n(t^\star(x)\,u | x,y).$$
The convergence results investigated in the sequel are established under an ergodicity condition ensured by the following assumption.

\begin{hypothese}\label{hyp:R}For any $(x,y)\in\mathcal{E}^2$, $R(\{y\}|x)>0$.\end{hypothese}

\begin{rem}\label{rm:ergo}
Under Assumption \ref{hyp:R}, the Markov chain $(Z_n)_{n\geq0}$ is irreducible, recurrent, and thus admits a unique invariant measure (up to a multiplicative constant) $\nu_\infty$ satisfying the eigenvalue problem,
$$\forall\,y\in\mathcal{E},~\sum_{x\in\mathcal{E}}\nu_\infty(\{x\}) R(\{y\}|x) = \nu_\infty(\{y\}).$$
The Markov chain $(Z_n)_{n\geq0}$ satisfies therefore the almost sure ergodic theorem presented for example in \cite[Theorem 4.3.15]{DD12}.
\end{rem}

\noindent
First we deal with the question of the asymptotic behavior of the coefficients $\widehat{\theta}^n_p$ and $\widehat{R}^n$ appearing in \eqref{eq:def:estimateurlambda} under the unique additional condition imposed in Assumption \ref{hyp:R}. It should be remarked that the asymptotic properties of $\widehat{R}^n$ are well-known in particular in the ergodic framework imposed by Assumption \ref{hyp:R}. In particular, one may state \cite[4.4 Statistics of Markov Chains]{DD12} that, for any $(x,y)\in\mathcal{E}^2$, when $n$ goes to infinity,
\begin{equation}\label{eq:R:ass}
\widehat{R}^n(\{y\}|x)\stackrel{a.s.}{\longrightarrow} R(\{y\}|x))\qquad\text{and}\qquad\sqrt{n}\left(\widehat{R}^n(\{y\}|x))-R(\{y\}|x))\right) \stackrel{d}{\longrightarrow}\mathcal{N}\left(0,\sigma^2_{R}(x,y))\right),
\end{equation}
where the asymptotic variance $\sigma^2_R(x,y)$ is defined by the following formula,
$$
\sigma^2_R(x,y) = \frac{R(\{y\}|x)(1-R(\{y\}|x))}{\nu_\infty(\{x\})}.
$$
The next result states the consistency and the asymptotic normality of $\widehat{\theta}^n_p$, $p\geq0$. For the sake of readability, $\widetilde{G}$ denotes the conditional survival function associated with the rate $\widetilde{\lambda}$,
$$\forall\,(x,y)\in\mathcal{E}^2,~\forall\,0\leq t<t^\star(x),~\widetilde{G}(t|x,y)=\exp\left[-\widetilde{\Lambda}(t|x,y)\right],$$
where the cumulative rate $\widetilde{\Lambda}$ has already been defined in \eqref{eq:Lambdatilde}.
\begin{prop1}
\label{prop:theta}
Under Assumptions \ref{hyp:lambdatilde}, \ref{hyp:tstar} and \ref{hyp:R}, for any couple $(x,y)\in\mathcal{E}^2$, when $n$ goes to infinity,
$$\widehat{\theta}^n_p(x,y)\stackrel{\mathbb{P}}{\longrightarrow} \theta_p(x,y)\qquad\text{and}\qquad\sqrt{n}\left(\widehat{\theta}^n_p(x,y)-\theta_p(x,y)\right) \stackrel{d}{\longrightarrow}\mathcal{N}\left(0,\sigma^2_{\theta_p}(x,y)\right),$$
where the asymptotic variance $\sigma^2_{\theta_p}(x,y)$ is defined by
$$\sigma^2_{\theta_p}(x,y) = \frac{1}{R(\{y\}|x)\nu_\infty(\{x\})t^\star(x)}\int_0^1\frac{B_p(s)^2\widetilde{\lambda}(t^\star(x)s|x,y)}{\widetilde{G}(t^\star(x)s|x,y)}\ud s.$$
\end{prop1}

\noindent
The asymptotic normality established in Proposition \ref{prop:theta} is of first importance in the estimation procedure presented in this paper. Indeed, one may propose a procedure to test the nullity of coefficients $\theta_p(x,y)$ thanks to this result. The unknown variance $\sigma^2_{\theta_p}(x,y)$ may be easily estimated by
\begin{equation}
\label{eq:estim:var}
\widehat{\sigma_{\theta_p}^2}^n(x,y)=\frac{1}{\widehat{R}^n(\{y\}|x)\widehat{\nu}^n_\infty(\{x\})t^\star(x)^2}\int_0^1B_p(s)^2\,\ud\!\!\left[\widehat{\widetilde{G}}^n(t^\star(x)s|x,y)^{-1}\right],
\end{equation}
where $\widehat{\nu}^n_\infty(\{x\})=n^{-1}\sum_{i=0}^{n-1}\mathbb{1}_{\{Z_i=x\}}$ intuitively estimates the invariant probability $\nu_\infty(\{x\})$ and, for any $0\leq s<t^\star(x)$, $\widehat{\widetilde{G}}^n(s|x,y)$ is a Fleming-Harrington type estimator of the conditional survival function $\widetilde{G}(s|x,y)$,
$$\widehat{\widetilde{G}}^n(s|x,y) = \exp\left[-\widehat{\widetilde{\Lambda}}^n(s|x,y)\right].$$
The procedure is based on the next result that provides the asymptotic behavior of the test statistic $T^n_{p}(x,y)$ defined by
$$T^n_{p}(x,y) = \frac{n \widehat{\theta}_p^n(x,y)^2}{\widehat{\sigma_{\theta_p}^2}^n(x,y)}$$
under the hypotheses $\theta_p(x,y)=0$ and $\theta_p(x,y)\neq0$.

\begin{cor}\label{prop:test}
Let $p\geq0$. Assume that $B_p$ is continuously differentiable on $[0,1]$. Under Assumptions \ref{hyp:lambdatilde}, \ref{hyp:tstar} and \ref{hyp:R}, for any couple $(x,y)\in\mathcal{E}^2$, under the null hypothesis $\theta_p(x,y)=0$, when $n$ goes to infinity,
$$T^n_{p}(x,y)\stackrel{d}{\longrightarrow} \chi^2(1),$$
where $\chi^2(1)$ denotes the chi-squared distribution with $1$ degree of freedom. Under the alternative hypothesis $\theta_p\neq0$, when $n$ goes to infinity, the test statistic $T^n_{p}(x,y)$ goes to infinity.
\end{cor}

\noindent
The test procedure resulting from the preceding corollary may be defined as follows. If the test statistic $T^n_{p}(x,y)$ is greater than $q_\alpha$ where $q_\alpha$ is defined by $\prob(\chi^2(1)\leq q_\alpha)=1-\alpha$, then one rejects the null hypothesis $\theta_p=0$. In the other case, one accepts the null hypothesis $\theta_p=0$.

\medskip

\noindent
From now on, we focus on the estimation of the jump function $\lambda$ from estimates $\widehat{R}^n$ and $\widehat{\theta}_p^n$ under some additional conditions.

\begin{hypotheses}
\label{hyps:bpsupp}
There exists a sequence $(\tau_n)_{n\geq0}$ such that $\gamma_n=o(\sqrt{n})$ where $(\gamma_n)_{n\geq0}$ is defined by,
$$\forall\,n\geq0,~\gamma_n=\sup_{s\in[0,1]}\left|\sum_{p=0}^{\tau_n}B_p(0)B_p(s)\right|.$$
In addition, we assume that,
$$\forall\,(x,y)\in\mathcal{E}^2,~\forall\,0\leq s\leq1,~\lim_{n\to\infty}\sum_{p=\tau_{n+1}}^{\infty} \langle B_p,\widetilde{\lambda}(\cdot\,t^\star(x)|x,y)\rangle B_p(s) =0.$$
\end{hypotheses}

\begin{rem}\label{rem:taun}
Conditions imposed in Assumptions \ref{hyps:bpsupp} are of course fulfilled when $(B_p)_{1\leq p\leq P}$ is a finite orthonormal family of continuous functions on $[0,1]$, such as a family of orthonormal splines. In this case, we can always choose $\tau_n=P$ and set $B_p=0$ for $p\geq P+1$. But these conditions are also satisfied when $(B_p)_{p\geq0}$ is the Fourier basis of $\mathbb{L}^2_{[0,1]}$. Notice that in this case, due to the presence of $B_p(0)$ in the definition of $\gamma_n$, only the cosine terms remain and we have, for any choice of $\tau_n$,
$$
\gamma_n=\sup_{s\in[0,1]}\left|1+\sqrt{2}\sum_{p=1}^{\tau_n}\cos(2\pi p s)\right|\leq 1+\sqrt{2}\tau_n.
$$
Assumptions \ref{hyps:bpsupp} are thus fulfilled as soon as $\tau_n=o(\sqrt{n})$. The same is true for the Legendre basis of $\mathbb{L}^2_{[0,1]}$.
\end{rem}

\noindent
In this setting, the rate $\lambda$ is estimated by,
\begin{equation}\label{eq:def:estimateurlambda}
\forall\,x\in\mathcal{E},~\widehat{\lambda}^n(x) = \sum_{p=0}^{\tau_n} B_p(0) \sum_{y\in\mathcal{E}}\widehat{R}^n(\{y\}|x) \widehat{\theta}^n_p(x,y).
\end{equation}
Our main result of convergence is stated in the following proposition.
\begin{prop1}\label{cor}
Under Assumptions \ref{hyp:lambdatilde}, \ref{hyp:tstar}, \ref{hyp:R} and \ref{hyps:bpsupp}, when $n$ goes to infinity,
$$\sup_{x\in\mathcal{E}}\left|\widehat{\lambda}^n(x) - \lambda(x)\right|\stackrel{\mathbb{P}}{\longrightarrow} 0.$$
\end{prop1}
\noindent
The number $\tau_n$ of terms used in the estimation \eqref{eq:def:estimateurlambda} of $\lambda$ is a tuning parameter that should be chosen in a judicious manner in practical situations. The resulting statistical analysis is much less complex than the methodology provided in \cite{AzaisSJS14} which requires to tune both two discretization steps and a bandwidth parameter. In addition, we would like to emphasize that Remark \ref{eq:def:estimateurlambda} provides a crucial information on this question since the assumptions imposed in the paper are satisfied whenever $\tau_n=o(\sqrt{n})$.

\section{Numerical illustration}
\label{s:4}

In this short simulation study, we consider a variant with discrete transitions of the well-known TCP process (see for example \cite[4.1 The TCP window-size process]{ABGKZ2014} and the references therein). The TCP window-size process appears as a scaling limit of the transmission rate of a server uploading packets on the Internet according to the algorithm used in the TCP (Transmission Control Protocol) in order to mitigate congestions. Usually, this process is defined on $\mathbb{R}_+$ but we consider here that the state space is only $[0,1]$. In other words, there may be some forced jumps when the path reaches the boundary. In this model, the deterministic motion $\Phi$ is defined by,
$$\forall\,x\in[0,1],~\forall\,t\geq0,~\Phi(t|x) = x+t.$$
As a consequence, the deterministic exit time $t^\star$ satisfies, for any $x\in[0,1]$, $t^\star(x)=1-x$. Usually, the transition kernel of this model is given, for any $x\in\mathbb{R}_+$, by $Q^c(\cdot|x) = \delta_{\{x/2\}}$. We propose a discrete version of this transition kernel that satisfies the constraints imposed in the paper. The support of $Q$ is the set $\mathcal{E}$ defined by
$$\mathcal{E}=\left\{0,\frac{1}{N},\dots,\frac{N-1}{N}\right\},$$
where $N$ is some integer greater than $2$. From any $x\in[0,1]$, the theoretical proportions $Q(\{y\}|x)$, $y\in\mathcal{E}$, are defined by
$$Q(\{y\}|x) \propto \frac{1}{1+\sqrt[4]{|y-x/2|}}.$$
$Q$ is thus a discrete and irreducible approximation of the usual transition kernel $Q^c$. A typical trajectory of the process is presented in Figure \ref{fig:tcp}. The transition distribution $Q(\cdot|x)$, for $x=0.987$, is displayed in Figure \ref{fig:transition} for $N=10$.

\begin{minipage}[c]{.45\textwidth}
\includegraphics[width=8cm]{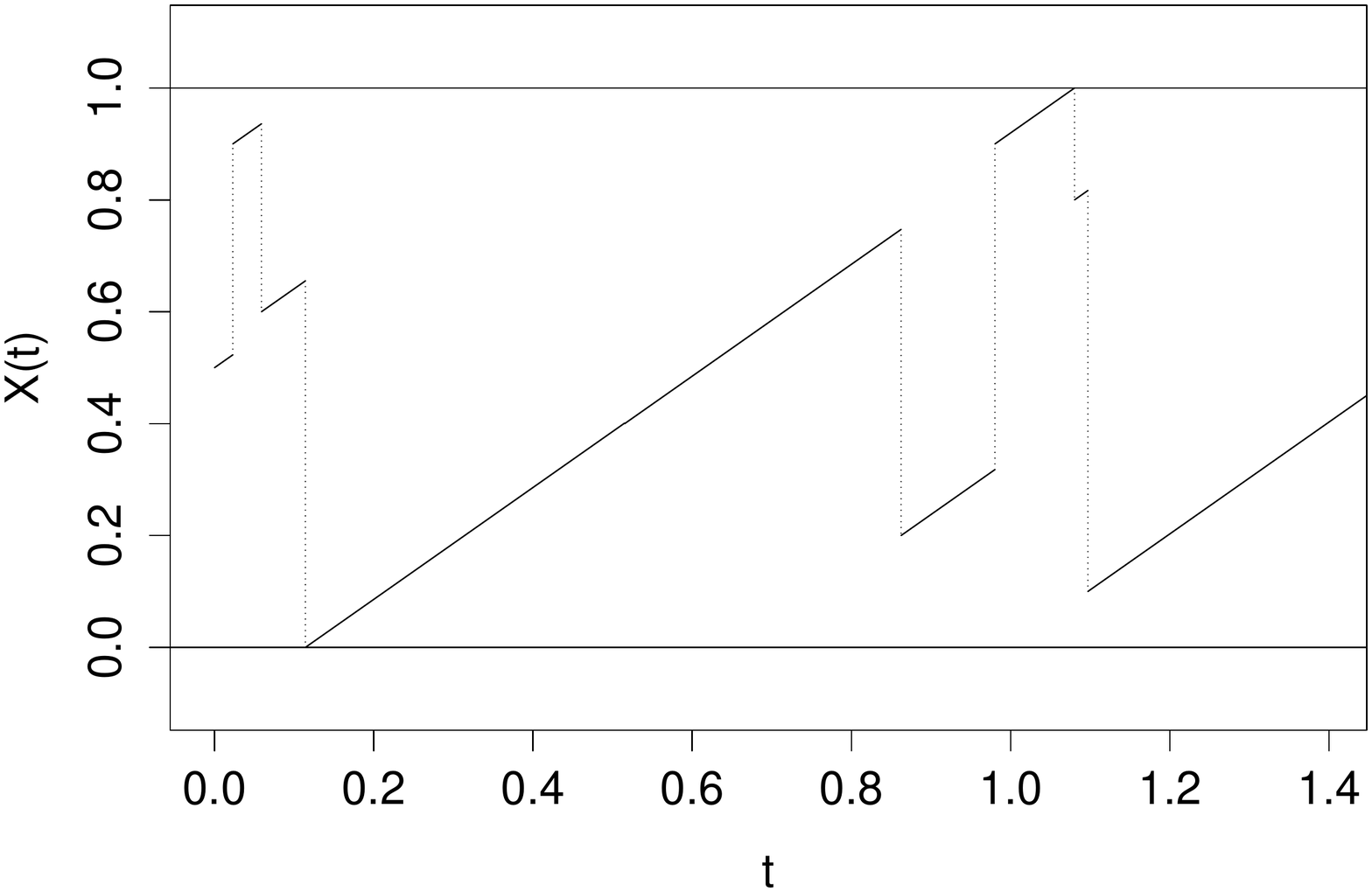}
\captionof{figure}{Typical trajectory of the variant with discrete transitions of the TCP window-size process with a forced jump between times $1$ and $1.2$.}
\label{fig:tcp}
\end{minipage}\qquad
\begin{minipage}[c]{.45\textwidth}
\includegraphics[width=8cm]{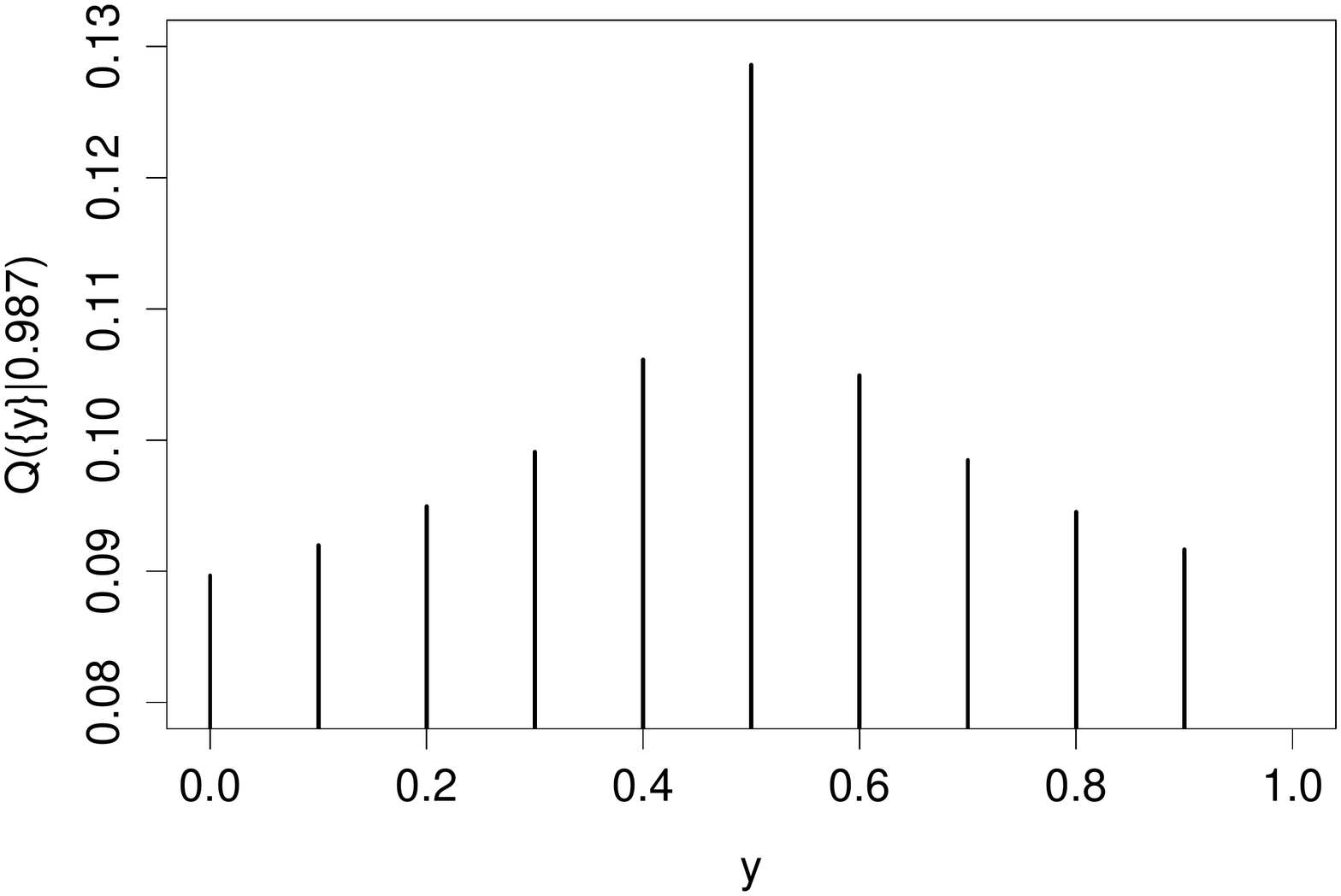}
\captionof{figure}{Transition distribution $Q(\cdot|x)$ on the $\mathcal{E}$ with $N=10$ elements for $x=0.987$. The highest probability stands for $y=1/2$, close to $x/2$.}
\label{fig:transition}
\end{minipage}

\vspace{0.5cm}

\begin{figure}[h]
\centering
\includegraphics[width=8cm]{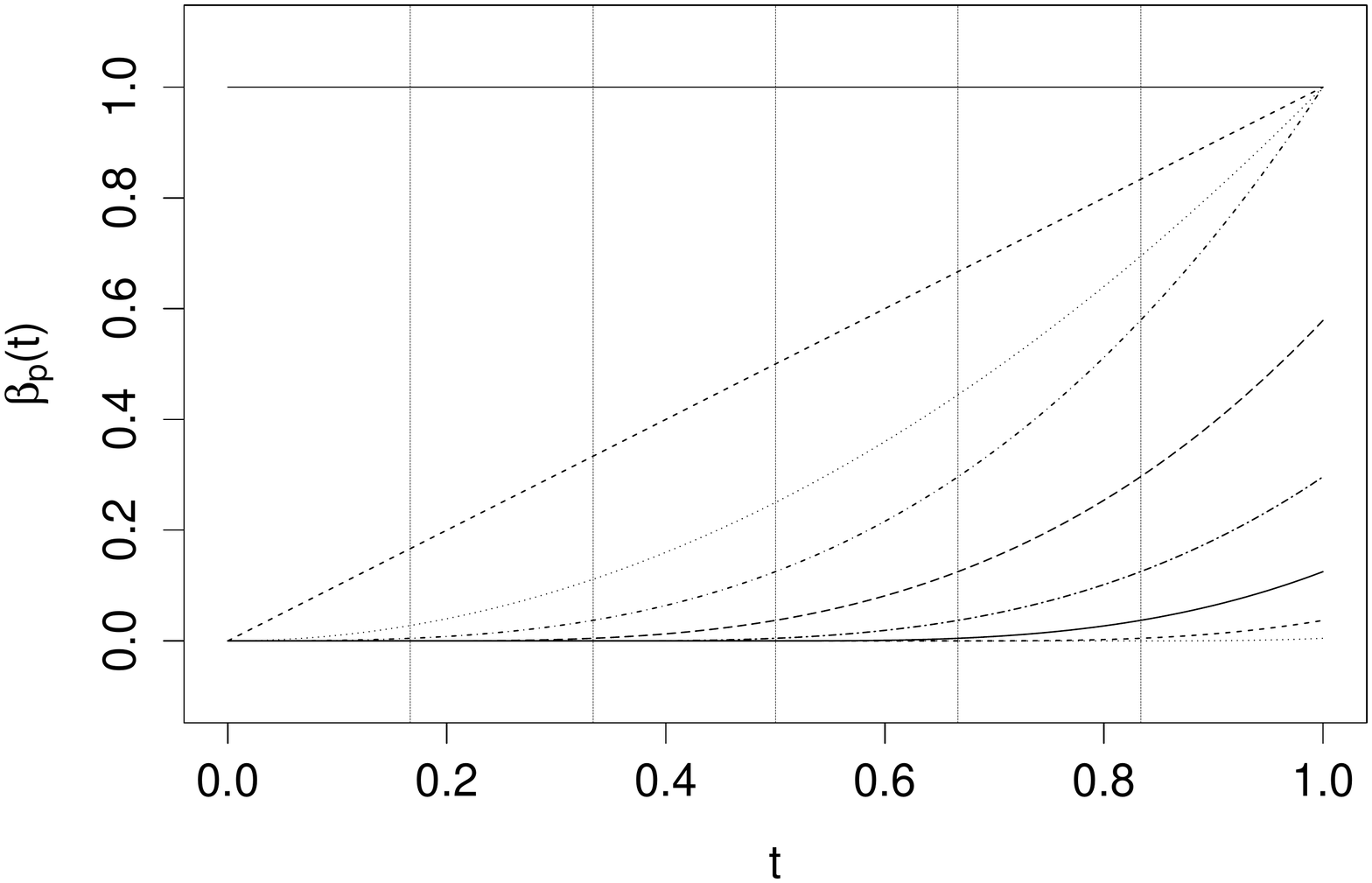}
\includegraphics[width=8cm]{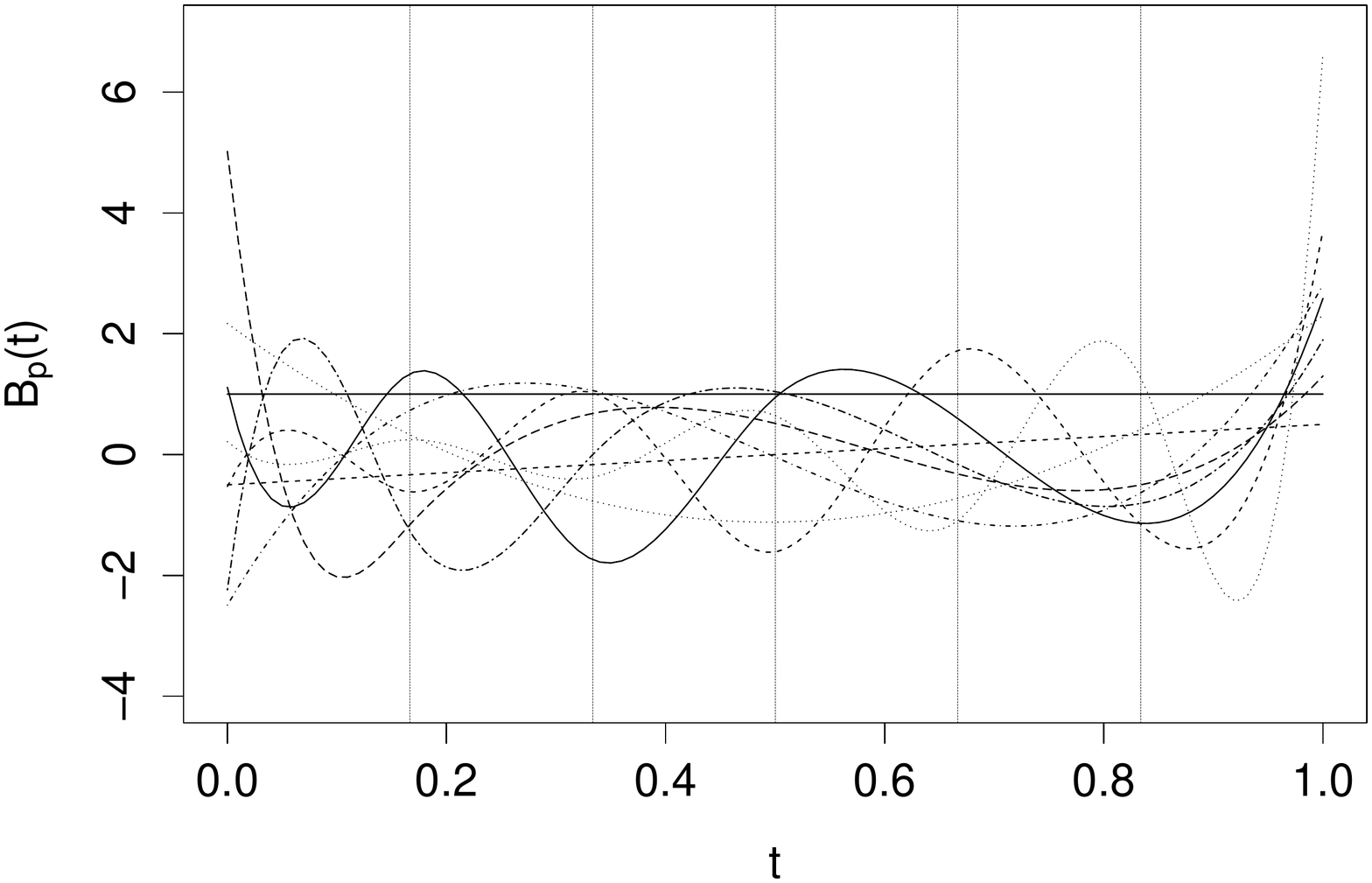}
\caption{Basis $(\beta_p)_{1\leq p\leq9}$ defined in \eqref{eq:basepason} of cubic splines with $5$ nodes uniformly distributed in the interval $[0,1]$ (left) and its orthonormal version obtained by using the Gram-Schmidt process (right).}
\label{fig:bases}
\end{figure}

\noindent
In our simulations, we propose to smooth the increments of the Nelson-Aalen estimator \eqref{eq:naestimate} in the functional space of cubic splines with 5 nodes at $\xi_k=k/6$, $1\leq k\leq 5$. The dimension of this functional space is thus $9$ and a basis is given by
\begin{equation}\label{eq:basepason}
(\beta_p)_{1\leq p\leq 9} = \left(\mathbb{I}\,;\,x\mapsto x\,;\,x\mapsto x^2\,;\,x\mapsto x^3\,;\,x\mapsto (x-k/6)_+^3~:~1\leq k\leq 5\right),
\end{equation}
where $\alpha_+=\max(\alpha,0)$ for any $\alpha\in\mathbb{R}$. Numerically, one may easily derive an orthonormal basis $(B_p)_{1\leq p\leq9}$ by using the Gram-Schmidt process.
We refer the reader to Figure \ref{fig:bases} for a graphical representation of the basis $(\beta_p)_{1\leq p\leq 9}$ and its orthonormal version $(B_p)_{1\leq p\leq9}$ obtained by the Gram-Schmidt algorithm.

\vspace{0.5cm}

\begin{figure}[h]
\centering
\includegraphics[width=8cm]{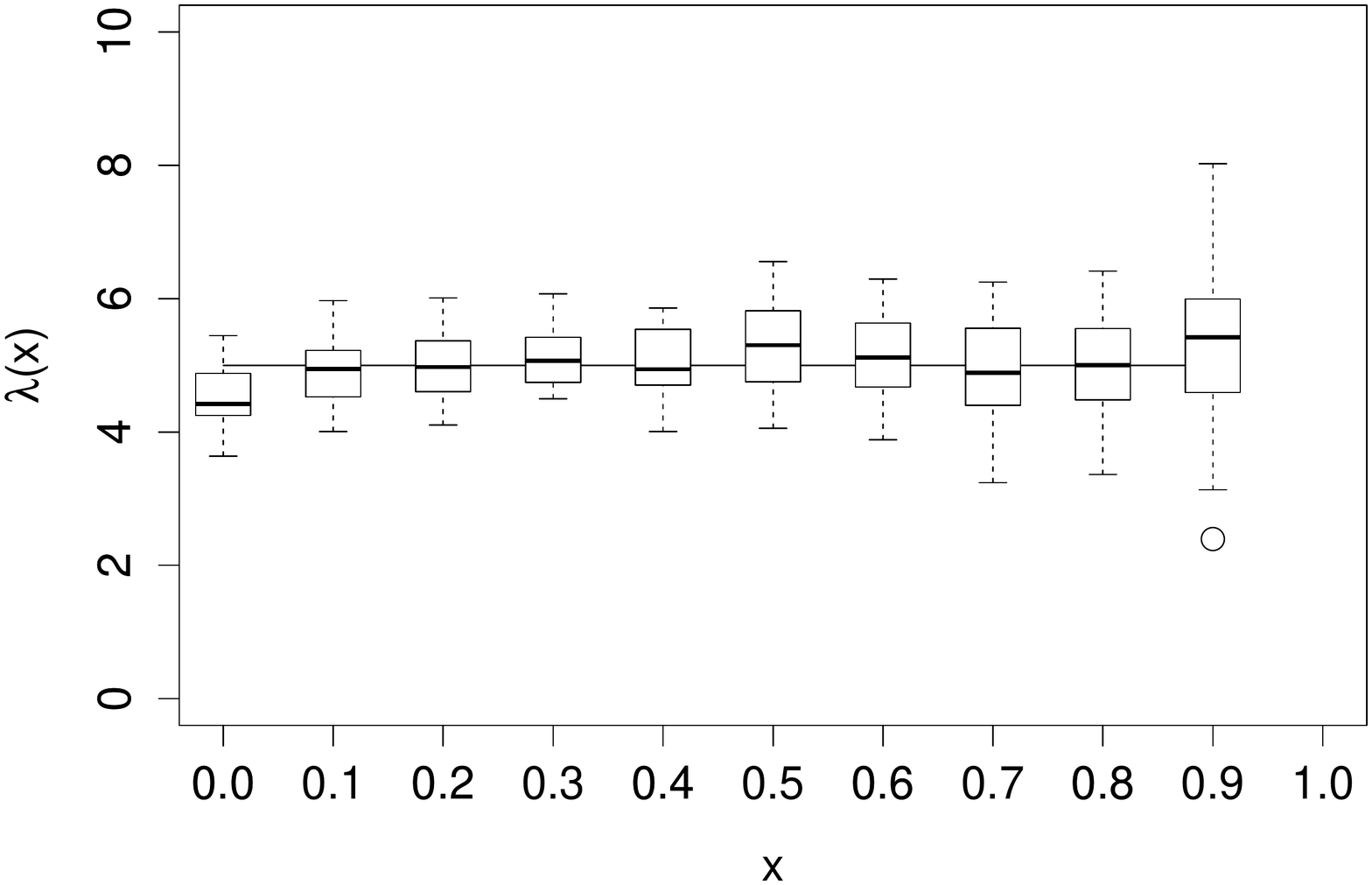} \includegraphics[width=8cm]{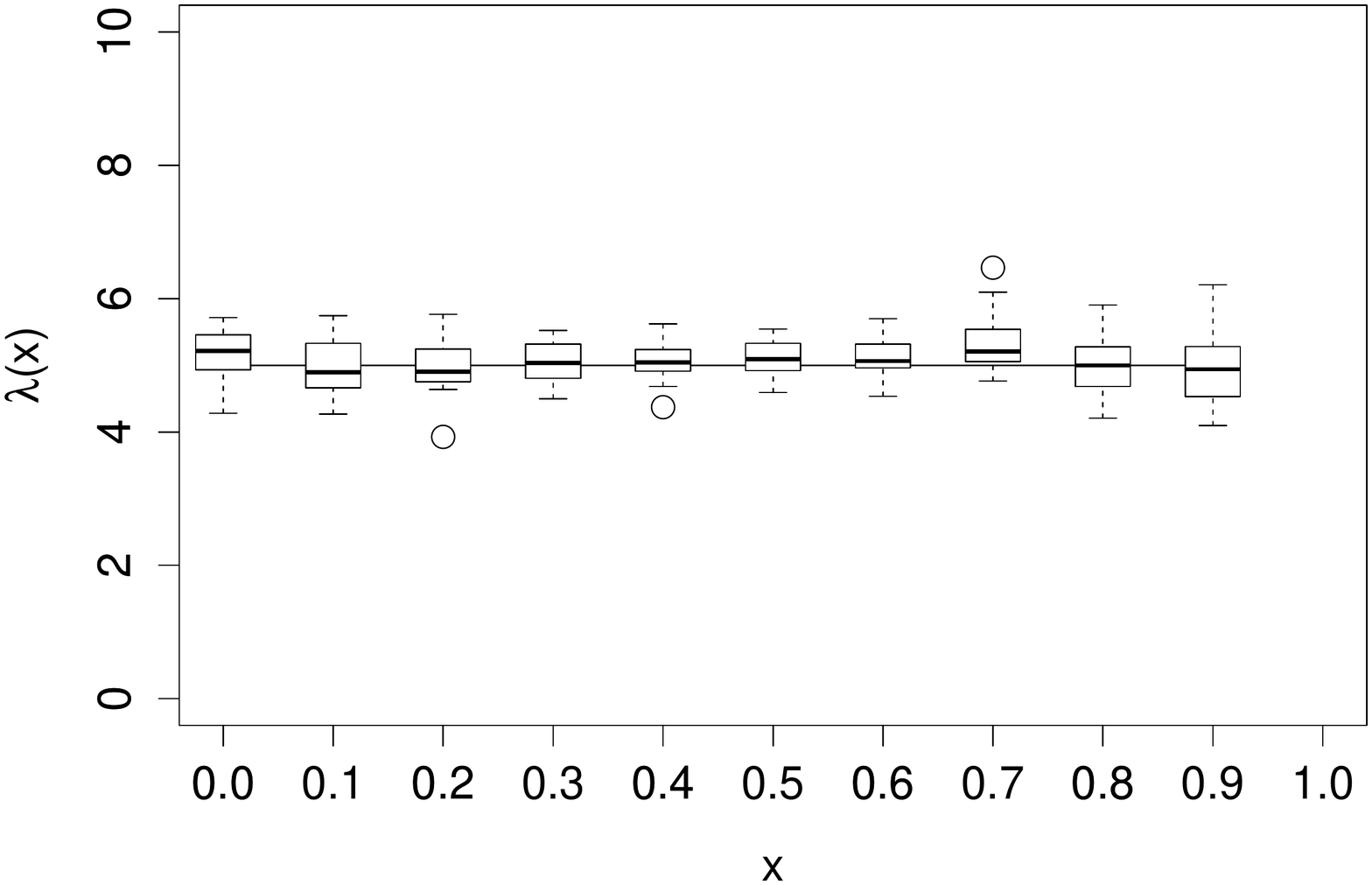}
\caption{Estimation of $\lambda(x)=5$ (full line) from $n=20\,000$ (left) and $n=50\,000$ (right) observed jumps. The boxplots have been computed over $100$ replicates.}
\label{fig:estim20000}
\end{figure}

\begin{figure}[h]
\centering
\includegraphics[width=8cm]{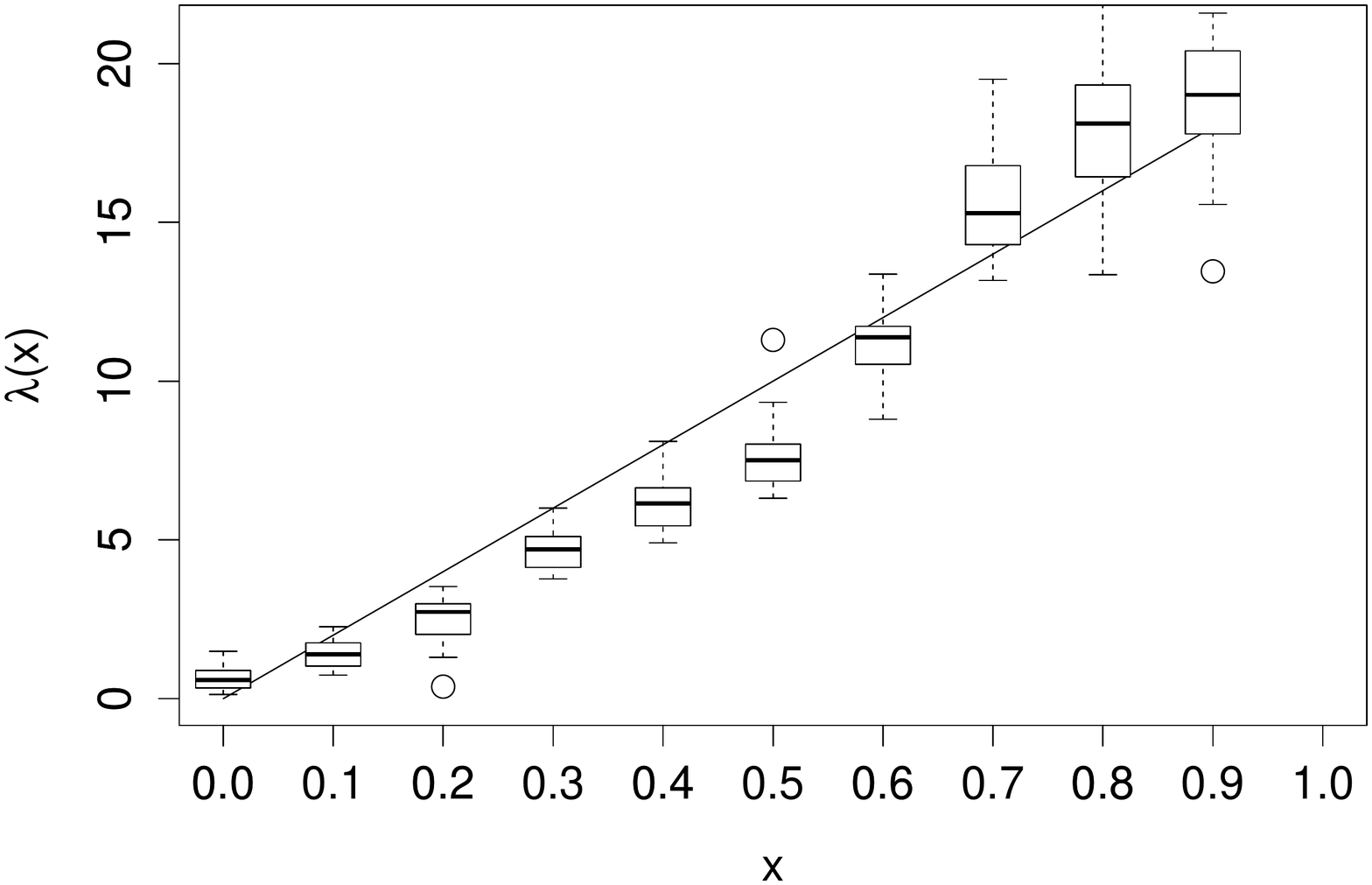} \includegraphics[width=8cm]{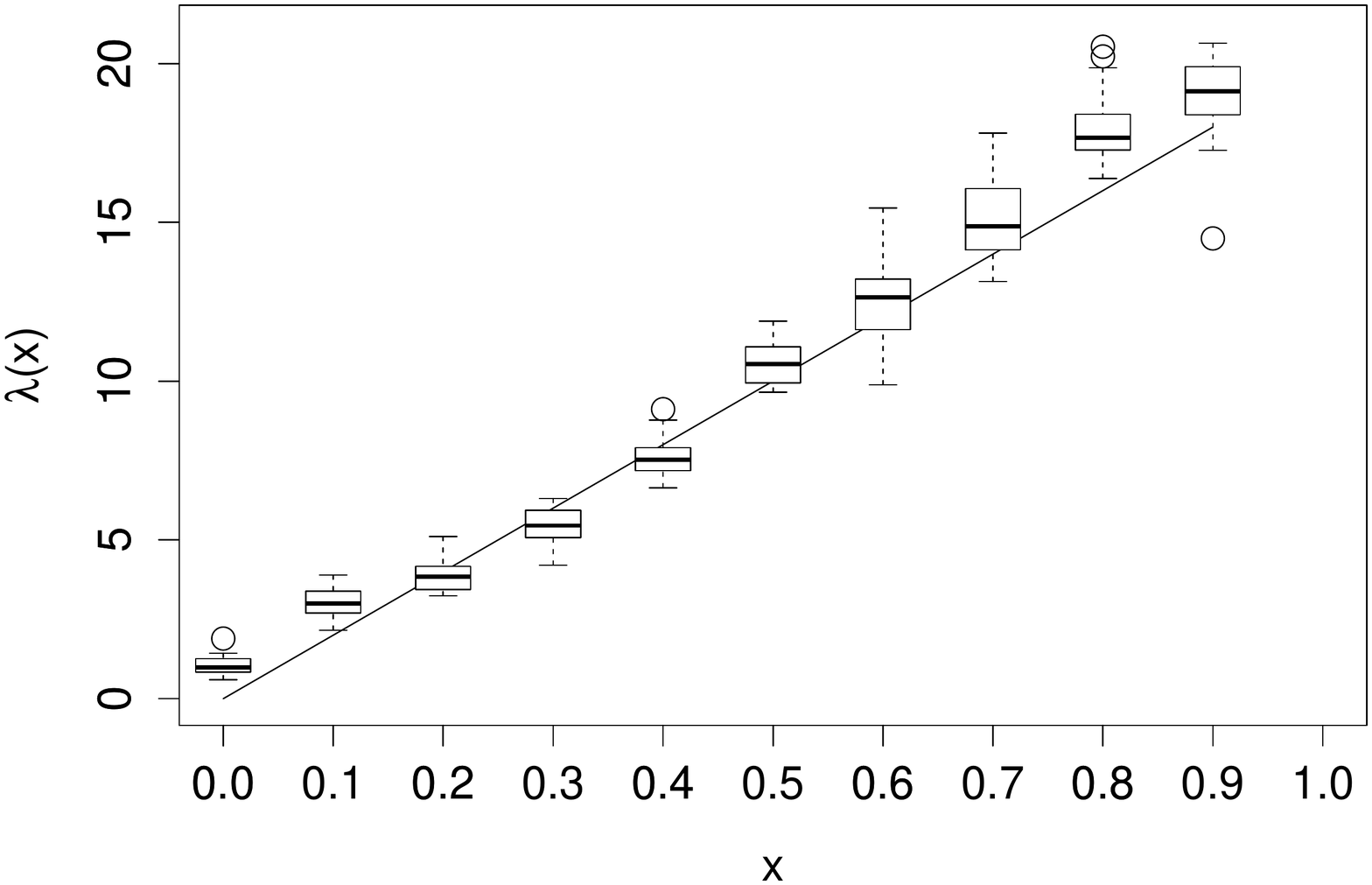}
\caption{Estimation of $\lambda(x)=20\,x$ (full line) from $n=20\,000$ (left) and $n=50\,000$ (right) observed jumps. The boxplots have been computed over $100$ replicates.}
\label{fig:estim50000}
\end{figure}

\noindent
We investigate the situations $\lambda(x)=5$ (exponential inter-jumping times right-censored by $1-x$) and $\lambda(x)=20\,x$ (Weibull inter-jumping times right-censored by $1-x$), $x\in[0,1]$. Estimation results computed from trajectories of $n=20\,000$ and $n=50\,000$ jumps are presented in Figures \ref{fig:estim20000} and \ref{fig:estim50000}. In both cases, we find out the true shape of the jump rate thanks to the estimation procedure. Bias and variance are small, in particular from $n=50\,000$ observed jumps. Therefore the methodology developed in this paper performs pretty well on this application example.

\appendix
\section{Proofs}
\label{s:app:proofs}

\subsection{Proof of Proposition \ref{prop:lambda}}
\label{ss:proof1}

By \eqref{eq:def:lambdatilde} and summing over all the possible values $y\in\mathcal{E}$, one has, for all $x\in\mathcal{E}$,
$$\forall\,0\leq t<t^\star(x),~ f(t|x) = \sum_{y\in\mathcal{E}} \widetilde{\lambda}(t|x,y) \left[\int_t^{t^\star(x)}\!\!\!f(s|x)Q(\{y\}|\Phi(s|x))\ud s + G(t^\star(x)|x)Q(\{y\}|\Phi(t^\star(x)|x))\right],$$
because $\sum_{y\in\mathcal{E}}Q(\{y\}|\Phi(x,t))=1$. Furthermore, in light of \eqref{eq:Glambdaf} together with $\Phi(0|x)=x$,
$$f(0|x) = \lambda(x)G(0|x) = \lambda(x),$$
since $G(0|x) = \prob(T_1>0|X(0)=x) = 1$ by \eqref{eq:G:S} and \eqref{eq:standard}. Thus,
$$\forall\,x\in\mathcal{E},~\lambda(x)=\sum_{y\in\mathcal{E}} \widetilde{\lambda}(0|x,y) \left[\int_0^{t^\star(x)}\!\!\!f(s|x)Q(\{y\}|\Phi(s|x))\ud s + G(t^\star(x)|x)Q(\{y\}|\Phi(t^\star(x)|x))\right].$$
Let us remark that, for any couple $(x,y)\in\mathcal{E}^2$, one has
\begin{eqnarray*}
\prob(Z_1=y|Z_0=x) &=& \prob(Z_1=y, S_1\in[0,t^\star(x))|Z_0=x) + \prob(Z_1=y,S_1=t^\star(x)|Z_0=x)\\
&=&\quad\int_0^{t^\star(x)}\prob(Z_1=y|S_1=t,Z_0=x)\prob(S_1\in\ud t|Z_0=x)\\
&&+~\prob(Z_1=y|S_1=t^\star(x),Z_0=x)\prob(S_1=t^\star(x)|Z_0=x)\\
&=&\int_0^{t^\star(x)}Q(\{y\}|\Phi(t|x) f(t|x)\ud t + Q(\{y\}|\Phi(t^\star(x)|x))G(t^\star(x)|x),
\end{eqnarray*}
by \eqref{eq:G:S} and because $f(\cdot|x)$ is the density of $S_1$ on the interval $[0,t^\star(x))$ conditionally on the event $\{Z_0=x\}$. As a consequence, one has,
$$\forall\,x\in\mathcal{E},~\lambda(x)=\sum_{y\in\mathcal{E}} \widetilde{\lambda}(0|x,y)R(\{y\}|x).$$
Together with Assumption \ref{hyp:lambdatilde}, this states the expected result.\hfill$\Box$


\subsection{Proof of Proposition \ref{prop:theta}}

\subsubsection{Convergence in probability}

Let $(x,y)$ be in $\mathcal{E}^2$ and $p\geq0$. The following decomposition holds,
\begin{eqnarray*}
\widehat{\theta}^n_p(x,y)-\theta_p(x,y)&=&\quad~\frac{1}{t^\star(x)}\int_0^1 Y^n(t^\star(x)s|x,y)^\ast B_p(s) \ud M^n(t^\star(x)s|x,y)\\
&&-~\int_0^1 B_p(s) \widetilde{\lambda}(t^\star(x)s|x,y)\mathbb{1}_{\{Y^n(t^\star(x)s|x,y)=0\}}\ud s.
\end{eqnarray*}
Moreover, for $0\leq t\leq1$, the term
$$
\widetilde{M}^n_p(t|x,y)=\frac{1}{t^\star(x)}\int_0^t Y^n(t^\star(x)s|x,y)^\ast B_p(s) \ud M^n(t^\star(x)s|x,y)
$$
defines a square integrable c\`adl\`ag martingale (with respect to some $t^\star(x)$-dependent filtration) with `angular brackets' process defined, for $0\leq t\leq1$, by
$$
\frac{1}{t^\star(x)}\int_0^t Y^n(t^\star(x)s|x,y)^\ast  B_p(s)^2\widetilde{\lambda}(t^\star(x)s|x,y)\ud s.
$$
The jumps of the martingale are given, for $0\leq t\leq1$, by
$$
\frac{1}{t^\star(x)}B_p(t) Y^n(t^\star(x)t|x,y)^\ast\Delta M^n(t^\star(x)t|x,y).
$$
Remark that the two above processes almost surely go to $0$ uniformly on $[0,1]$ since, by the almost sure ergodic theorem (see Remark \ref{rm:ergo}), $n^{-1}Y^n(t^\star(x)s|x,y)^\ast$ almost surely converges, when $n$ goes to infinity, towards
\begin{equation}\label{eq:ps}
\frac{1}{\mathbb{P}(S_1\geq t^\star(x)s,Z_1=y|Z_0=x)\nu_\infty(\{x\})}\geq\min_{x,y\in\cal E}\frac{1}{\mathbb{P}(Z_1=y|Z_0=x)\nu_\infty(\{x\})}>0.
\end{equation}
Then, using the fact that $Y^n(t^\star(x)\cdot|x,y)$ is a decreasing process, we have
\begin{eqnarray*}
\left|\widehat{\theta}^n_p(x,y)-\theta_p(x,y)\right|&\leq&\quad~\left|\frac{1}{t^\star(x)}\int_0^1 Y^n(t^\star(x)s|x,y)^\ast B_p(s) \ud M^n(t^\star(x)s|x,y)\right|\\
&&~+\,\big\|\widetilde{\lambda}(t^\star(x)\cdot|x,y)\big\|_2\,\mathbb{1}_{\{Y^n(t^\star(x)|x,y)=0\}}.
\end{eqnarray*}
By Assumption \ref{hyp:R}, the term $\mathbb{1}_{\{Y^n(t^\star(x)|x,y)=0\}}$ converges to $0$ in $\mathbb{L}^1$ and thus in probability. Using a Bernstein type inequality such as \cite[Corollary 3.4]{Berny}, we have that the martingale term converges also to $0$ in probability since the `angular brackets' process and the jumps almost surely go to $0$. This shows the result.\hfill$\Box$

\subsubsection{Central limit theorem}

We now proceed to the proof of the associated central limit theorem. At first, notice that, almost surely, using dominated convergence (in $s$, recalling that $Y^n(t^\star(x)\cdot|x,y)$ is a decreasing process for the domination),
\begin{align*}
\lim_{n\to\infty}\,n\int_0^1 Y^n(t^\star(x)s|x,y)^\ast  B_p(s)^2\widetilde{\lambda}(t^\star(x)s|x,y)\ud s
&=\int_0^1 \frac{B_p(s)^2\widetilde{\lambda}(t^\star(x)s|x,y)}{\mathbb{P}(S_1\geq t^\star(x)s,Z_1=y|Z_0=x)\nu_\infty(\{x\})}\ud s\\
&=\,t^\star(x) \sigma^2_{\theta_p}(x,y).
\end{align*}
In the same manner, for any $\e>0$, we almost surely have
$$
\lim_{n\to\infty}\,\frac{n}{t^\star(x)}\int_0^1 Y^n(t^\star(x)s|x,y)^\ast  B_p(s)^2\widetilde{\lambda}(t^\star(x)s|x,y)\mathbb{1}_{\left\{\frac{\sqrt n}{t^\star(x)}B_p(s) Y^n(t^\star(x)s|x,y)^\ast\Delta M^n(t^\star(x)s|x,y)>\e\right\}}\ud s=0.
$$
We can thus apply \cite[Theorem V.1]{Rebo} to obtain that the process
$$
\frac{1}{t^\star(x)}\int_0^1 Y^n(t^\star(x)s|x,y)^\ast B_p(s) \ud M^n(t^\star(x)s|x,y)
$$
converges in law towards a centered Gaussian random variable with variance $\sigma^2_{\theta_p}(x,y)$. We also have, almost surely,
$$
\sqrt{n}\int_0^1 B_p(s) \widetilde{\lambda}(t^\star(x)s|x,y)\mathbb{1}_{\{Y^n(s|x,y)=0\}}\ud s~\leq~\int_0^1 B_p(s) \widetilde{\lambda}(t^\star(x)s|x,y)\ud s \sqrt{n}\mathbb{1}_{\{Y^n(1|x,y)=0\}}.
$$
Since $\mathbb{1}_{\{Y^n(1|x,y)=0\}}$ belongs to $\{0,1\}$ and almost surely converges to $0$, we also have that $\sqrt{n}\mathbb{1}_{\{Y^n(1|x,y)=0\}}$ almost surely converges to $0$ and thus, by comparison, almost surely,
$$
\lim_{n\to\infty}\,\sqrt{n}\int_0^1 B_p(s) \widetilde{\lambda}(t^\star(x)s|x,y)\mathbb{1}_{\{Y^n(t^\star(x)s|x,y)=0\}}\ud s=0.
$$
An application of Slutsky's theorem implies that the process
$$
\sqrt{n}\left(\widehat{\theta}^n_p(x,y)-\theta_p(x,y)\right)
$$
converges in law towards a centered Gaussian random variable with variance $\sigma^2_{\theta_p}(x,y)$.\hfill$\Box$


\subsection{Proof of Corollary \ref{prop:test}}

We are going to show that the estimated variance \eqref{eq:estim:var} converges in probability towards $\sigma_{\theta_p}^2(x,y)$. At first, by the almost sure ergodic theorem (see Remark \ref{rm:ergo} and \eqref{eq:R:ass}), we have
$$
\frac{1}{\widehat{R}^n(\{y\}|x)\widehat{\nu}^n_\infty(\{x\})t^\star(x)^2}\stackrel{a.s.}{\longrightarrow}\frac{1}{R(\{y\}|x)\nu_\infty(\{x\})t^\star(x)^2}.
$$
As a consequence, by Slutsky's theorem, it remains to show that
$$
\int_0^1B_p(s)^2\,\ud\!\!\left[\widehat{\widetilde{G}}^n(t^\star(x)s|x,y)^{-1}\right]
$$
converges in probability towards
$$
\int_0^1B_p(s)^2\,\ud\!\!\left[\widetilde{G}(t^\star(x)s|x,y)^{-1}\right].
$$
Recall that $B^2_p$ is a continuous function of finite variation and $\widehat{\widetilde{G}}^n(t^\star(x)s|x,y)^{-1}$ is a semi-martingale. Thus, by integration by parts, we have
\begin{eqnarray*}
\int_0^1B_p(s)^2\,\ud\!\!\left[\widehat{\widetilde{G}}^n(t^\star(x)s|x,y)^{-1}\right]&=&\quad~\int_0^1B_p(s)^2\,\ud\!\!\left[\widetilde{G}(t^\star(x)s|x,y)^{-1}\right]\\
&&+~B_p(1)^2\left[\widehat{\widetilde{G}}^n(t^\star(x)|x,y)^{-1}-\widetilde{G}(t^\star(x)|x,y)^{-1}\right]\\
&&-~B_p(0)^2\left[\widehat{\widetilde{G}}^n(0|x,y)^{-1}-\widetilde{G}(0|x,y)^{-1}\right]\\
&&-~\int_0^1\left[\widehat{\widetilde{G}}^n(t^\star(x)s|x,y)^{-1}-\widetilde{G}(t^\star(x)s|x,y)^{-1}\right]\,\ud B_p(s)^2.
\end{eqnarray*}
Note that
$$
\widehat{\widetilde{G}}^n(0|x,y)^{-1}=\widetilde{G}(0|x,y)^{-1}=1.
$$
Let us show that the term
$$
\int_0^1\left[\widehat{\widetilde{G}}^n(t^\star(x)s|x,y)^{-1}-\widetilde{G}(t^\star(x)s|x,y)^{-1}\right]\,\ud B_p(s)^2
$$
goes to $0$ in probability. We have, for any $\delta>0$,
\begin{align*}
\mathbb{P}&\left(\left|\int_0^1\left[\widehat{\widetilde{G}}^n(t^\star(x)s|x,y)^{-1}-\widetilde{G}(t^\star(x)s|x,y)^{-1}\right]\,\ud B_p(s)^2\right|\geq\delta\right)\\
&\leq~\mathbb{P}\left(\sup_{s\in[0,1]}\left|\widehat{\widetilde{G}}^n(t^\star(x)s|x,y)^{-1}-\widetilde{G}(t^\star(x)s|x,y)^{-1}\right| V^1_0(B_p^2)\geq \delta\right)
\end{align*}
where $V^1_0(B_p^2)$ is the total variation of $B_p^2$ over $[0,1]$, which is finite since $B_p$ is continuously differentiable on the interval $[0,1]$ by assumption. Then,\begin{align*}
&\mathbb{P}\left(\sup_{s\in[0,1]}\left|\widehat{\widetilde{G}}^n(t^\star(x)s|x,y)^{-1}-\widetilde{G}(t^\star(x)s|x,y)^{-1}\right| V^1_0(B_p^2)\geq \delta\right)\\
&=~\mathbb{P}\left(\sup_{s\in[0,1]}\left|\exp\left[\widehat{\widetilde{\Lambda}}^n(t^\star(x)s|x,y)\right]-\exp\left[{\widetilde{\Lambda}(t^\star(x)s|x,y)}\right]\right| V^1_0(B_p^2)\geq \delta\right)\\
&\leq~\mathbb{P}\Bigg(\left\{\sup_{s\in[0,1]}\left|\exp\left[\widehat{\widetilde{\Lambda}}^n(t^\star(x)s|x,y)\right]-\exp\left[{\widetilde{\Lambda}(t^\star(x)s|x,y)}\right]\right| V^1_0(B_p^2)\geq \delta\right\}
\\
&\quad\quad\quad~~\cap~\left\{\sup_{s\in[0,1]}\left|\widehat{\widetilde{\Lambda}}^n(t^\star(x)s|x,y)-\widetilde{\Lambda}(t^\star(x)s|x,y)\right|<\delta\right\}\Bigg)\\
&\quad~+\mathbb{P}\left(\sup_{s\in[0,1]}\left|\widehat{\widetilde{\Lambda}}^n(t^\star(x)s|x,y)-\widetilde{\Lambda}(t^\star(x)s|x,y)\right|\geq\delta\right)\\
&\leq~\mathbb{P}\Bigg(\left\{\exp\left[{\delta+\widetilde{\Lambda}(t^\star(x)|x,y)}\right]\sup_{s\in[0,1]}\left|\widehat{\widetilde{\Lambda}}^n(t^\star(x)s|x,y)-\widetilde{\Lambda}(t^\star(x)s|x,y)\right| V^1_0(B_p^2)\geq \delta\right\}\\
&\quad\quad\quad~~\cap~\left\{\sup_{s\in[0,1]}\left|\widehat{\widetilde{\Lambda}}^n(t^\star(x)s|x,y)-\widetilde{\Lambda}(t^\star(x)s|x,y)\right|<\delta\right\}\Bigg)\\
&\quad~+\mathbb{P}\left(\sup_{s\in[0,1]}\left|\widehat{\widetilde{\Lambda}}^n(t^\star(x)s|x,y)-\widetilde{\Lambda}(t^\star(x)s|x,y)\right|\geq\delta\right)
\end{align*}
by the mean value theorem. Therefore, we have
\begin{align*}
\mathbb{P}&\left(\sup_{s\in[0,1]}\left|\widehat{\widetilde{G}}^n(t^\star(x)s|x,y)^{-1}-\widetilde{G}(t^\star(x)s|x,y)^{-1}\right| V^1_0(B_p^2)\geq \delta\right)\\
&\leq~\mathbb{P}\left(\exp\left[{\widetilde{\Lambda}(t^\star(x)|x,y)}\right]\sup_{s\in[0,1]}\left|\widehat{\widetilde{\Lambda}}^n(t^\star(x)s|x,y)-\widetilde{\Lambda}(t^\star(x)s|x,y)\right| V^1_0(B_p^2)\geq \delta\exp(-\delta)\right)\\
&\quad~+\mathbb{P}\left(\sup_{s\in[0,1]}\left|\widehat{\widetilde{\Lambda}}^n(t^\star(x)s|x,y)-\widetilde{\Lambda}(t^\star(x)s|x,y)\right|\geq\delta\right).
\end{align*}
This latter term goes to $0$ since $\widehat{\widetilde{\Lambda}}^n(\cdot|x,y)$ converges to $\widetilde{\Lambda}(\cdot|x,y)$ uniformly in probability in light of \cite[Proposition 3.7]{AzaisSJS14}. As a consequence, we have proven that, when $n$ goes to infinity,
$$
\widehat{\sigma_{\theta_p}^2}^n(x,y)\stackrel{\mathbb{P}}{\longrightarrow}\sigma_{\theta_p}^2(x,y).
$$
The rest of the proof relies on Slutsky's theorem again and standard calculus.\hfill$\Box$


\subsection{Proof of Proposition \ref{cor}}

Let $\delta>0$ and $x\in\mathcal{E}$. We begin the proof with some steps of elementary simplifications. One may write
$$
\widehat{\lambda}^n(x) - \lambda(x)=\widehat{\lambda}^n(x) -\lambda^n(x)+\lambda^n(x)- \lambda(x),
$$
where $\lambda^n(x)$ is defined by
$$
\lambda^n(x)=\sum_{p=0}^{\tau_n} B_p(0) \sum_{y\in\mathcal{E}}R(\{y\}|x) \theta_p(x,y).
$$
Then,
$$
\mathbb{P}\left(\sup_{x\in\mathcal{E}}\left|\widehat{\lambda}^n(x) - \lambda(x)\right|\geq\delta\right) \leq \mathbb{P}\left(\sup_{x\in\mathcal{E}}\left|\widehat{\lambda}^n(x) - \lambda^n(x)\right|\geq\frac\delta2\right)+\mathbb{P}\left(\sup_{x\in\mathcal{E}}\left|\lambda^n(x) - \lambda(x)\right|\geq\frac\delta2\right).
$$
By virtue of Assumptions \ref{hyps:bpsupp}, the second term vanishes for $n$ large enough. For the first term, the state space $\cal E$ being finite we have
$$
\mathbb{P}\left(\sup_{x\in\mathcal{E}}\left|\widehat{\lambda}^n(x) - \lambda^n(x)\right|\geq\frac\delta2\right) \leq \sum_{x\in{\cal E}}\mathbb{P}\left(\left|\widehat{\lambda}^n(x) - \lambda^n(x)\right|\geq\frac{\delta}{2\#\cal E}\right) .
$$
Now, for $x\in\cal E$ held fixed, we have
\begin{align*}
\mathbb{P}&\left(\left|\widehat{\lambda}^n(x) - \lambda^n(x)\right|\geq\frac{\delta}{2\#\cal E}\right)\\
&=~\mathbb{P}\left(\Bigg|\sum_{p=0}^{\tau_n} B_p(0) \sum_{y\in\mathcal{E}}\widehat{R}^n(\{y\}|x) \widehat{\theta}^n_p(x,y)-\sum_{p=0}^{\tau_n} B_p(0) \sum_{y\in\mathcal{E}}R(\{y\}|x) \theta_p(x,y)\Bigg|\geq\frac{\delta}{2\#\cal E}\right)\\
&\leq~\sum_{y\in\mathcal{E}}\mathbb{P}\left(\left|\sum_{p=0}^{\tau_n} B_p(0) \widehat{R}^n(\{y\}|x) \widehat{\theta}^n_p(x,y)-\sum_{p=0}^{\tau_n} B_p(0) R(\{y\}|x) \theta_p(x,y)\right|\geq\frac{\delta}{2\#{\cal E}^2}\right).
\end{align*}
We thus work from now on with fixed $x,y\in\cal E$. We have
\begin{align*}
\mathbb{P}&\left(\left|\sum_{p=0}^{\tau_n}B_p(0) \widehat{R}^n(\{y\}|x) \widehat{\theta}^n_p(x,y)-\sum_{p=0}^{\tau_n} B_p(0) R(\{y\}|x) \theta_p(x,y)\right|\geq\frac{\delta}{2\#{\cal E}^2}\right)\\
&=~\mathbb{P}\left(\left|\widehat{R}^n(\{y\}|x)\sum_{p=0}^{\tau_n} B_p(0) \left[\widehat{\theta}^n_p(x,y)-\theta_p(x,y)\right]-\left[R(\{y\}|x)-\widehat{R}^n(\{y\}|x)\right]\sum_{p=0}^{\tau_n} B_p(0)  \theta_p(x,y)\right|\geq\frac{\delta}{2\#{\cal E}^2}\right)\\
&\leq~\mathbb{P}\left(\left|\widehat{R}^n(\{y\}|x)\sum_{p=0}^{\tau_n} B_p(0)\left[\widehat{\theta}^n_p(x,y)-\theta_p(x,y)\right]\right|\geq\frac{\delta}{4\#{\cal E}^2}\right)\\
&\quad~+\mathbb{P}\left(\left|\left[R(\{y\}|x)-\widehat{R}^n(\{y\}|x)\right]\sum_{p=0}^{\tau_n} B_p(0)  \theta_p(x,y)\right|\geq\frac{\delta}{4\#{\cal E}^2}\right).
\end{align*}
Remark that, by Assumptions \ref{hyps:bpsupp},
$$
\left|\sum_{p=0}^{\tau_n} B_p(0)  \theta_p(x,y)\right|=\left|\sum_{p=0}^{\tau_n} B_p(0)  \int_0^1\widetilde{\lambda}\left(t^\star(x)\,u|x,y\right) B_p(u)\ud u\right|\leq\kappa_1\gamma_n<\infty,
$$
for some deterministic constant $\kappa_1$. Thus, for $n$ large enough, we have
$$
\mathbb{P}\left(\left| \left[R(\{y\}|x)-\widehat{R}^n(\{y\}|x)\right]\sum_{p=0}^{\tau_n}B_p(0)  \theta_p(x,y)\right|\geq\frac{\delta}{4\#{\cal E}^2}\right)\leq\mathbb{P}\left(\gamma_n\left| R(\{y\}|x)-\widehat{R}^n(\{y\}|x)\right|\geq\frac{\delta}{4\kappa_1\#{\cal E}^2}\right).
$$
This latter term may be written as
$$
\mathbb{P}\left(\gamma_n\left| R(\{y\}|x)-\widehat{R}^n(\{y\}|x)\right|\geq\frac{\delta}{4\kappa_1\#{\cal E}^2}\right)=\mathbb{P}\left(\frac{\gamma_n}{\sqrt{n}}\sqrt{n}\left| R(\{y\}|x)-\widehat{R}^n(\{y\}|x)\right|\geq\frac{\delta}{4\kappa_1\#{\cal E}^2}\right).
$$
Notice that, according to \eqref{eq:R:ass}, the sequence $\sqrt{n}[R(\{y\}|x)-\widehat{R}^n(\{y\}|x)]$ converges in law towards some Gaussian random variable. Then, since $\gamma_n/\sqrt{n}$ goes to $0$ by Assumptions \ref{hyps:bpsupp}, using Slutsky's theorem, the sequence $\gamma_n[R(\{y\}|x)-\widehat{R}^n(\{y\}|x)]$ goes to $0$ in distribution. This implies that
$$
\lim_{n\to\infty}\,\mathbb{P}\left(\gamma_n\left| R(\{y\}|x)-\widehat{R}^n(\{y\}|x)\right|\geq\frac{\delta}{4\kappa_1\#{\cal E}^2}\right)=0.
$$
The remaining term
$$
\mathbb{P}\left(\left|\widehat{R}^n(\{y\}|x)\sum_{p=0}^{\tau_n} B_p(0)\left[\widehat{\theta}^n_p(x,y)-\theta_p(x,y)\right]\right|\geq\frac{\delta}{4\#{\cal E}^2}\right)
$$
requires more attention and will be treated with similar tools as the ones used in the proof of Proposition \ref{prop:theta}. First, $\widehat{R}^n(\{y\}|x)$ being positive and bounded by one, we have
$$
\mathbb{P}\left(\left|\widehat{R}^n(\{y\}|x)\sum_{p=0}^{\tau_n} B_p(0)\left[\widehat{\theta}^n_p(x,y)-\theta_p(x,y)\right]\right|\geq\frac{\delta}{4\#{\cal E}^2}\right)\leq \mathbb{P}\left(\left|\sum_{p=0}^{\tau_n} B_p(0) \left[\widehat{\theta}^n_p(x,y)-\theta_p(x,y)\right]\right|\geq\frac{\delta}{4\#{\cal E}^2}\right).
$$
For any $0\leq p\leq \tau_n$, we have
\begin{eqnarray*}
B_p(0)\widehat{\theta}^n_p(x,y)-B_p(0)\theta_p(x,y) &=&\quad~\frac{1}{t^\star(x)}\int_0^1 Y^n(t^\star(x)s|x,y)^\ast B_p(0)B_p(s) \ud M^n(t^\star(x)s|x,y)\\
&&-~\int_0^1 B_p(0)B_p(s) \widetilde{\lambda}(t^\star(x)s|x,y)\mathbb{1}_{\{Y^n(t^\star(x)s|x,y)=0\}}\ud s.
\end{eqnarray*}
For $0\leq t\leq1$, the term
$$
\widetilde{M}^n_p(t|x,y)=\frac{1}{t^\star(x)}\int_0^t Y^n(t^\star(x)s|x,y)^\ast B_p(0)B_p(s) \ud M^n(t^\star(x)s|x,y)
$$
defines a square integrable c\`adl\`ag martingale with respect to some filtration. Moreover, the process defined for $0\leq t\leq1$ by $\mu^n(t|x,y)=\sum_{p=0}^{\tau_n}\widetilde{M}^n_p$ is still a square integrable c\`adl\`ag martingale. Its `angular brackets' process is defined, for $0\leq t\leq1$, by
$$
\langle\mu^n(\cdot|x,y)\rangle_t=\frac{1}{t^\star(x)}\int_0^t Y^n(t^\star(x)s|x,y)^\ast \left(\sum_{p=0}^{\tau_n}B_p(0) B_p(s)\right)^2\widetilde{\lambda}(t^\star(x)s|x,y)\ud s.
$$
In addition, the jumps of this martingale are given, for $0\leq t\leq1$, by
$$
\Delta\mu^n(t|x,y)=\frac{1}{t^\star(x)}\sum_{p=0}^{\tau_n}B_p(0) B_p(t) Y^n(t^\star(x)t|x,y)^\ast\Delta M^n(t^\star(x)t|x,y).
$$
We have
\begin{align*}
\mathbb{P}&\left(\left|\sum_{p=0}^{\tau_n} B_p(0)  [\widehat{\theta}^n_p(x,y)-\theta_p(x,y)]\right|\geq\frac{\delta}{4\#{\cal E}^2}\right)\\
&\leq~\mathbb{P}\left(\left|\mu^n(1,x,y)\right|\geq\frac{\delta}{8\#{\cal E}^2}\right)+\mathbb{P}\left(\left|\sum_{p=0}^{\tau_n}\int_0^1 B_p(0)B_p(s) \widetilde{\lambda}(t^\star(x)s|x,y)\mathbb{1}_{\{Y^n(t^\star(x)s|x,y)=0\}}\ud s\right|\geq\frac{\delta}{8\#{\cal E}^2}\right).
\end{align*}
As highlighted before, $n^{-1}Y^n(t^\star(x)s|x,y)^\ast$ almost surely converges towards a deterministic value $\kappa_2>0$. As a consequence, we have that, almost surely, the time spent by $Y^n(t^\star(x)\cdot|x,y)$ in $0$ is null for large $n$, and
$$
\lim_{n\to\infty}\,\mathbb{P}\left(\left|\sum_{p=0}^{\tau_n}\int_0^1 B_p(0)B_p(s) \widetilde{\lambda}(t^\star(x)s|x,y)\mathbb{1}_{\{Y^n(t^\star(x)s|x,y)=0\}}\ud s\right|\geq\frac{\delta}{8\#{\cal E}^2}\right)=0.
$$
Writing $c_1=t^\star(x)^{-1}\big\|\widetilde{\lambda}(t^\star(x)\cdot|x,y)\big\|_2(\kappa_2+1)$ and  using \cite[Corollary 3.4]{Berny}, we have, for any $a>0$,
\begin{eqnarray*}
\mathbb{P}\left(\left|\mu^n(1|x,y)\right|\geq\frac{\delta}{8\#{\cal E}^2}\right)&\leq &\quad2\exp\left[{-\frac12\left(\frac{\delta}{8\#{\cal E}^2}\right)^2\frac{n}{c_1\gamma^2_n}\psi\left(\frac{a\delta n}{8\#{\cal E}^2c_1\gamma^2_n}\right)}\right]+\mathbb{P}\left(\langle\mu^n(\cdot|x,y)\rangle_1> c_1\frac{\gamma^2_n}{n}\right)\\
&&+~\mathbb{P}\left(\sup_{s\in[0,1]}|\Delta\mu^n(s|x,y)|> a\right),
\end{eqnarray*}
where $\psi:[0,\infty)\to\mathbb{R}$ is defined by,
$$\forall\,x\geq0,~\psi(x) = \frac{2}{x^2} \int_0^x \log(1+y)\ud y.$$
Then, using \eqref{eq:ps} and noticing that the jumps of $M^n(\cdot|x,y)$ are bounded by one, this is not hard to see that, almost surely, for $n$ large enough, under Assumptions \ref{hyp:R} and \ref{hyps:bpsupp}, we have
$$
\sup_{s\in[0,1]}|\Delta\mu^n(s|x,y)|\,\leq\,\frac a2\,<\,a ,$$
and
$$\langle\mu^n(\cdot|x,y)\rangle_1\,\leq\, t^\star(x)^{-1}\big\|\widetilde{\lambda}(t^\star(x)\cdot|x,y)\big\|_2 \left(\kappa_2+\frac12\right)\frac{\gamma^2_n}{n}\,<\,c_1\frac{\gamma^2_n}{n}.
$$
Thus
$$
\lim_{n\to\infty}\,\mathbb{P}\left(\langle\mu^n(\cdot|x,y)\rangle_1> c_1\frac{\gamma^2_n}{n}\right)+\mathbb{P}\left(\sup_{s\in[0,1]}|\Delta\mu^n(s|x,y)|> a\right)=0.
$$
Since $\psi$ is bounded on $[0,\infty)$, we also have, thanks to Assumptions \ref{hyps:bpsupp},
$$
\lim_{n\to\infty}\,2\exp\left[{-\frac12\delta^2\frac{n}{\gamma^2_n}c_1\psi\left(ac_1\delta\frac{n}{\gamma^2_n}\right)}\right]=0.
$$
Thus,
$$
\lim_{n\to\infty}\,\mathbb{P}\left(\left|\mu^n(1|x,y)\right|\geq\frac{\delta}{8\#{\cal E}^2}\right)=0.
$$
The set $\cal E$ being finite, the result follows by summing over $x$ and $y$.\hfill$\Box$

\bibliographystyle{plain}
\bibliography{main}

\end{document}